\documentclass{aastex631}

\usepackage{graphicx}
\usepackage{caption}
\usepackage{subcaption}
\begin{document}

\title{Exploring Dimensionality Reduction of SDSS Spectral Abundances}

\author{Qianyu Fan}
\affiliation{Department of Statistical Sciences, University of Toronto}

\begin{abstract}
High-resolution stellar spectra offer valuable insights into atmospheric parameters and chemical compositions. However, their inherent complexity and high-dimensionality present challenges in fully utilizing the information they contain. In this study, we utilize data from the Apache Point Observatory Galactic Evolution Experiment (APOGEE) within the Sloan Digital Sky Survey IV (SDSS-IV) to explore latent representations of chemical abundances by applying five dimensionality reduction techniques: PCA, t-SNE, UMAP, Autoencoder, and VAE. Through this exploration, we evaluate the preservation of information and compare reconstructed outputs with the original 19 chemical abundance data. Our findings reveal a performance ranking of PCA \textless{} UMAP \textless{} t-SNE \textless{} VAE \textless{} Autoencoder, through comparing their explained variance under optimized MSE. The performance of non-linear (Autoencoder and VAE) algorithms has approximately 10\% improvement compared to linear (PCA) algorithm. This difference can be referred to as the "non-linearity gap." Future work should focus on incorporating measurement errors into extension VAEs, thereby enhancing the reliability and interpretability of chemical abundance exploration in astronomical spectra.
\end{abstract}


\section{Introduction}  \label{sec:intro}
Stellar spectra play a vital role in revealing the fundamental properties of stars, including atmospheric parameters and chemical compositions, shedding light on the formation and evolution of galaxies within the Milky Way. The availability of an enormous amount of high-resolution stellar spectra today is attributed to many past and ongoing extensive spectroscopic surveys. Among these, the Apache Point Observatory Galactic Evolution Experiment (APOGEE) within the Sloan Digital Sky Survey IV (SDSS-IV) stands out for its comprehensive collection of high-resolution stellar spectra. APOGEE \citep{majewski2017apogee}  has amassed spectra from hundreds of thousands of stars in the inner Galaxy, leveraging near-infrared observations at high resolution ($R = 22,500$) and high signal-to-noise ratio ($S/N \geq 100$); these provide high-precision kinematical properties and chemical information.

High-resolution stellar spectra contain a wealth of information essential for understanding stellar properties and evolutionary processes. However, this information is often embedded in high-dimensional spaces where complex correlations exhibit non-linear patterns, posing significant challenges for comprehensive analysis. The issue of high dimensionality is not unique to stellar spectra but is pervasive across various domains including biomedical, web, education, medicine, business, and social media \citep{DRotherdomian}. In genomics, for instance, gene expression data collected from microarray experiments typically involve thousands of genes, each representing a dimension in the dataset. Similarly, in neuroscience, brain imaging techniques generate high-dimensional data representing neural activity across multiple regions of the brain.

These high-dimensional datasets present challenges in visualization, interpretation, and analysis. The curse of dimensionality exacerbates computational complexity and hinders the exploration of underlying structures and patterns. Consequently, dimensionality reduction techniques have emerged as indispensable tools for navigating high-dimensional spaces and extracting meaningful information.

There are many different dimensionality reduction techniques, including linear techniques and non-linear ones. In \cite{14DR} research, it reviewed 14 dimensionality reduction techniques: the linear techniques considered were Principal Component Analysis (PCA), Linear Discriminant Analysis (LDA), Singular Value Decomposition (SVD), Latent Semantic Analysis (LSA), Locality Preserving Projections (LPP), Independent Component Analysis (ICA) and Project Pursuit (PP); and non-linear techniques considered were Kernel Principal Component Analysis (KPCA), Multi-dimensional Scaling (MDS), Isomap, Locally Linear Embedding (LLE), Self-Organizing Map (SOM), Latent Vector Quantization (LVQ), t-Stochastic neighbor embedding (t-SNE) and Uniform Manifold Approximation and Projection (UMAP). 

Among these, LDA always requires labels, SVD and LPP have high computational complexity, LSA is primarily designed for text data, ICA assumes independent data distributions, and PP can be inefficient for large-scale data. Though PCA also struggles with nonlinear data, it is widely used in astronomy \citep{PCAwideuse}. In contrast to the traditional linear techniques, the non-linear techniques have the ability to deal with complex non-linear data. In particular, the real-world data, such as astronomy data is typically formed a highly non-linear manifold. To assess the improvement brought by nonlinear techniques in dimensionality reduction, so we adopted PCA as a baseline for comparison. For non-linear techniques, t-SNE and UMAP were opted since they offer robust performance compared with others. To pursuing more accuracy, astronomers also used neural networks (NN), like Autoencoder and variational autoencoder (VAE) for dimensionality reduction \citep{NNDR}. They can effectively handle high-dimensional data, capturing intricate patterns and non-linear data representations. 

In our study, we implemented five dimensionality reduction techniques, ranging from linear to non-linear methods, to uncover underlying structures and patterns in 2-dimensional space. Our goal is to explore latent representations of chemical abundances, evaluate the preservation of information during the dimensionality reduction process, and compare reconstruction results with the original data.

The remainder of the paper is structured as follows. We describe data and data preprocessing in Section 2. Then in Section 3, we introduce dimensionality reduction techniques and neural network approach. Section 4 summarizes the results and Section 5 discusses these techniques and the next steps before concluding in Section 6.

\section{Data}  \label{sec:data}
The data used in our study is from APOGEE Data Release 17 (DR17), which encompasses 19 chemical abundances and stellar parameters for 370,060 stars. This data was obtained through automated analysis of spectra using APOGEE Stellar Parameter and Chemical Abundances Pipeline (ASPCAP) software \citep{garcía2016aspcap}. We removed poor quality stars by ASPCAPFLAG and STARFLAG bitmasks to mitigate potential quality warnings and also removed outliers and applied standard scale, resulting in a final dataset comprising 133,891 stars.

We focused on chemical abundances, which refer to the relative abundance of different chemical elements in the atmosphere of stars. To quantify the relative amounts of individual elements, it uses the abundance ratio as the logarithm of the ratio of two metallic elements in a star relative to their ratio in the sun. For instance, if a star has a [C/FE] value of -0.08, it means its carbon abundance is 83\% that of the Sun. We are using 19 dimensions related to chemical abundances, including the metallicity [FE/H] and other 18 element abundances.

\section{Methods} \label{sec:methods}
\subsection{Dimensionality Reduction Techniques}
Our study aims to investigate latent representations of chemical abundances in a 2-dimensional space. We utilized five dimensionality reduction techniques to achieve this goal.

\subsubsection{Principle Component Analysis}
Principle Component Analysis (PCA) \citep{pca1993} is one of the oldest and most widely used techniques which is an unsupervised linear dimension reduction technique. PCA identifies the directions (principal components) in which the data varies the most and projects the data onto these directions, thus reducing the dimensionality while preserving the most important information. The first principal component captures the direction of maximum variance in the data, and each subsequent component captures the remaining variance orthogonal to the previous ones. Since we want to reduce dimension into 2, it identifies two orthogonal directions that capture the maximum variance in the data. 

\subsubsection{t-Distributed Stochastic Neighbor Embedding}
t-Distributed Stochastic Neighbor Embedding (t-SNE) \citep{tsne2008} is a non-linear algorithm that finds similarities between data points in high-dimensional spaces by modeling pairwise similarities using a Gaussian distribution. Then, t-SNE aims to find a low-dimensional representation of the data points in such a way that the similarities between points in the high-dimensional space are preserved as closely as possible. Since some points in the high dimension will be compressed into one point in the low dimension due to a short tail in the distribution, it uses t-distribution to create low-dimensional space. The algorithm minimizes the divergence between the probability distributions of pairwise similarities in the high-dimensional and low-dimensional spaces using a gradient descent optimization approach. By iteratively adjusting the positions of data points in the low-dimensional space, t-SNE seeks to create a mapping that reveals the underlying structure and clusters within the data.

\subsubsection{Uniform Manifold Approximation and Projection}
Uniform Manifold Approximation and Projection (UMAP) \citep{umap2018} is similar to t-SNE but with some key differences. It constructed a weighted graph and built a fuzzy topological representation to approximate the manifold structure of the high-dimensional data and preserve both local and global structure. Then, it optimizes a low-dimensional representation of the data through stochastic gradient descent, efficiently adjusting the low-dimensional embedding to best capture the essential features of the data. Additionally, it utilizes a constraint, where each data point is connected to at least its nearest neighbors, to ensure the whole graph is connected. This feature distinguishes UMAP, enabling it to better preserve the global structure of the chemical abundances.

\subsubsection{Autoencoder}
Autoencoder \citep{autoencoder2020} is a type of unsupervised feedforward neural network \citep{NN}. It consists of two main components: an encoder and a decoder. The encoder takes the input data and compresses it into a lower-dimensional representation, typically called the latent space. This compressed representation contains the most important features of the input data, effectively reducing its dimensionality. The decoder then takes this compressed representation and attempts to reconstruct the original input data from it. The goal of the autoencoder is to learn an encoding-decoding process that minimizes the difference between the input and the reconstructed output. In other words, it aims to learn a compact representation of the data that retains as much relevant information as possible. Autoencoders can be trained using various optimization techniques, such as gradient descent, to minimize a reconstruction loss function, such as mean squared error. In our case, it takes 19-dimensional chemical abundances as input and compresses it to produce a 2-dimensional latent representation, followed by a decoder that takes the latent representation and decompresses it to produce a reconstruction of the original data.

\subsubsection{Variational Autoencoder}
Variational Autoencoder (VAE) \citep{vae2013} is an extension of Autoencoder that encodes input as a distribution over the latent space rather than a single point, and the distribution is typically represented by the mean and log variance (used to guarantee variance is positive) parameters of a Gaussian distribution. This is achieved by adding a regularization term to the loss function, which is typically the KL divergence \citep{kldiv1951} between the distribution output by the encoder and the standard Gaussian distribution. To train the VAE using backpropagation, we need to compute gradients with respect to the parameters of the encoder and decoder networks. However, since the encoder outputs a probability distribution, we cannot directly backpropagate through sampling operations. The reparameterization trick addresses this issue by decoupling the sampling process from the parameters of the distribution. Instead of sampling directly from the distribution, we sample from a standard Gaussian distribution and then transform the samples using the parameters of the distribution output by the encoder.

\subsection{Statistical Measure}
To evaluate the preservation of information during dimensionality reduction, we applied explained variance as a metric to measure the performance of these techniques. Explained variance denotes the proportion of the total variance in the original 19-dimensional data that is captured in its 2-dimensional representation. 

\[\text{Explained Variance} = 1 - \frac{\text{Unexplained Variation}}{\text{Total Variation}}\] 

Among this formula, the total variance of 19-dimensional data can be computed by the trace of the variance-covariance matrix.

\[\text{trace(S)} = \sum_{j=1}^{m} S^2 = \sum_{j=1}^{m} (\frac{1}{n}\sum_{i=1}^{n}(X_{ij} - \bar{X}_{mn,ij})^2)\] where $m$ represents the number of dimensions, i.e., 19 dimensions, $n$ represents the number of objects, ${X}_{ij}$ denotes the value of the $j$-th dimension for the $i$-th object, and $\bar{X}_{mn,ij}$ denotes the mean value of the $j$-th dimension across all objects.

Unexplained variation refers to the portion of the total variability in the data that remains unaccounted, while Mean Squared Error (MSE) captures the difference between predicted reconstructed outputs and original data. They are related to each other, and so we can use MSE as a measure of unexplained variation.

\subsection{Neural Network Architectures}
\subsubsection{A Fully-connected Neural Network for t-SNE and UMAP}
While PCA provides inherent explained variance, we lack an explicit reconstruction function $f(Z_{i};\theta)$ for t-SNE and UMAP.

\[MSE(\theta) = \frac{1}{n}\sum_{i=1}^{n}(X_{i} - f(Z_{i};\theta)^2)\] where ${X}_{i}$ represents original data and ${Z}_{i}$ represents the data after dimension reduction.

To address this, we trained a fully-connected neural network to predict the reconstructed values for t-SNE and UMAP respectively. The neural network model was constructed using the PyTorch Sequential module, featuring a series of linear layers interspersed with Batch Normalization \citep{batchnorm2015} and LeakyReLU activation functions. The architecture in Figure 1 begins with a linear layer taking 2-dimensional input data (i.e., data after dimensionality reduction), which is then transformed into a 64-dimensional representation. This is followed by another linear layer reducing the dimension to 32, and finally, the output layer generates the expected 19-dimensional reconstructed output. Each linear transformation is followed by Batch Normalization to normalize the input and improve the stability and performance of the model by reducing internal covariate shift. The LeakyReLU activation is then applied to introduce non-linearity into the model and capture complex patterns in the data. The LeakyReLU activation function is defined as 

\[
\text{LeakyReLU}(x) = \left\{
\begin{array}{ll}
x & \text{for } x \geq 0 \\
ax & \text{for } x < 0, \text{ where } a = 0.01
\end{array}
\right.
\]
This introduces a small slope when input is negative, rather than setting the activation function to zero, helping to alleviate the vanishing gradient problem and enhance the model’s representational capacity.

We employed an Adam optimizer \citep{adam2015} with a learning rate of $1 \times 10^{-4}$ to minimize the Mean Squared Error (MSE) loss function, with hyperparameters tuned using the Optuna framework. The chemical abundance dataset consists of n objects across m dimensions, and the MSE is calculated as follows:

\[\text{MSE} = \frac{1}{m} \frac{1}{n} \sum_{i=1}^{n} \sum_{j=1}^{m} (X_{ij} - {X}_{mn,ij})^2\]
where $m$ represents the number of dimensions, i.e., 19 dimensions, $n$ represents the number of objects, ${X}_{ij}$ denotes the value of the $j$-th dimension for the $i$-th object, and ${X}_{mn,ij}$ denotes the predicted value of the $j$-th dimension across all objects.

The model was trained for $3,000$ epochs using mini-batches of data. Each epoch involved forward and backward passes, with model weights updated using the Adam optimizer. To ensure convergence, early stopping was implemented with a convergence threshold set to $1 \times 10^{-4}$. If the average loss on the test set showed no significant improvement for $5$ consecutive epochs, the model was considered to have converged. This mechanism effectively prevented overfitting by halting the training process when the model's performance on the test set no longer improves. Finally, the model with the lowest MSE on the test set was retained as the optimal model. 

\begin{figure}[ht]
    \centering
    \includegraphics[width=0.3\linewidth]{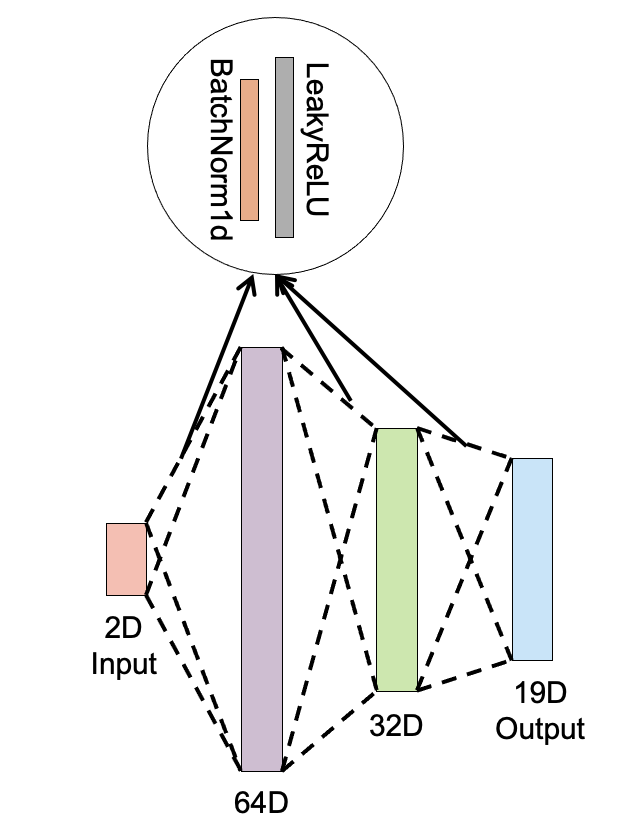}
    \caption{Structure of a fully-connected neural network with a 2-64-32-19 architecture. Each linear transformation is followed by Batch Normalization and LeakyReLU activation. We trained this neural network for t-SNE and UMAP respectively to predict the reconstructed values since they don't have explicit reconstruction function.Then, we use MSE to measure difference between reconstructed values and original data and calculate explained variance.}
    \label{fig:NN}
\end{figure}

\subsubsection{Autoencoder and VAE Neural Networks}
The architecture of both Autoencoder and VAE in Figure 2 is symmetric, with the encoder and decoder being mirror images of each other. To ensure consistency and facilitate comparative analysis with t-SNE and UMAP, both models utilized their decoder parts to reconstruct the 2-dimensional latent representations back to the original 19-dimensional space, so the decoder follows a 2-64-32-19 architecture. Given their symmetry, both models have a 19-32-64-2-64-32-19 architecture. 

\begin{figure}[ht]
    \centering
    \begin{subfigure}{0.4\textwidth}
        \includegraphics[width=\linewidth]{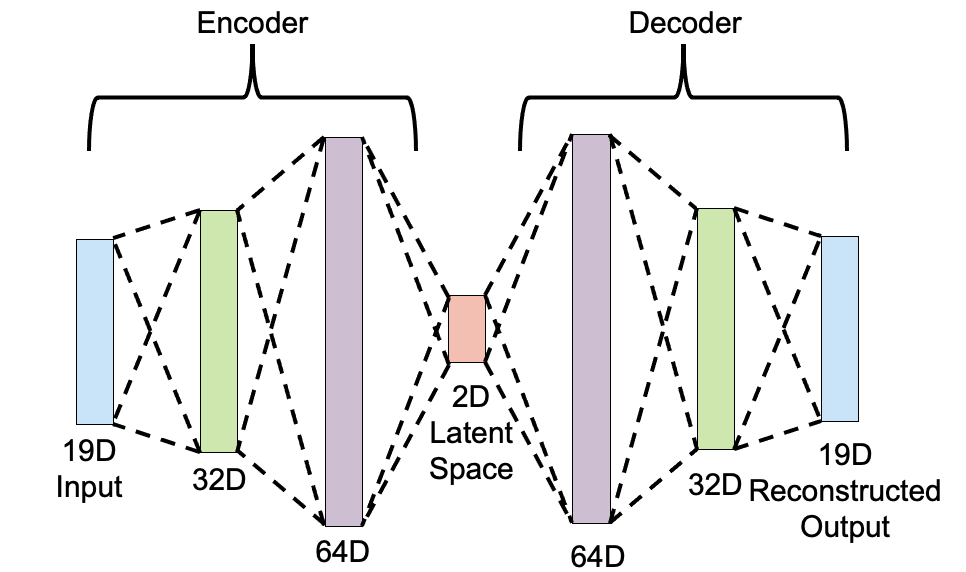}
        \caption{Structure of a autoencoder with a 19-32-64-2-64-32-19 architecture.}
        \label{fig:AE_archi}
    \end{subfigure}
 
    \medskip
    
    \begin{subfigure}{0.4\textwidth}
        \includegraphics[width=\linewidth]{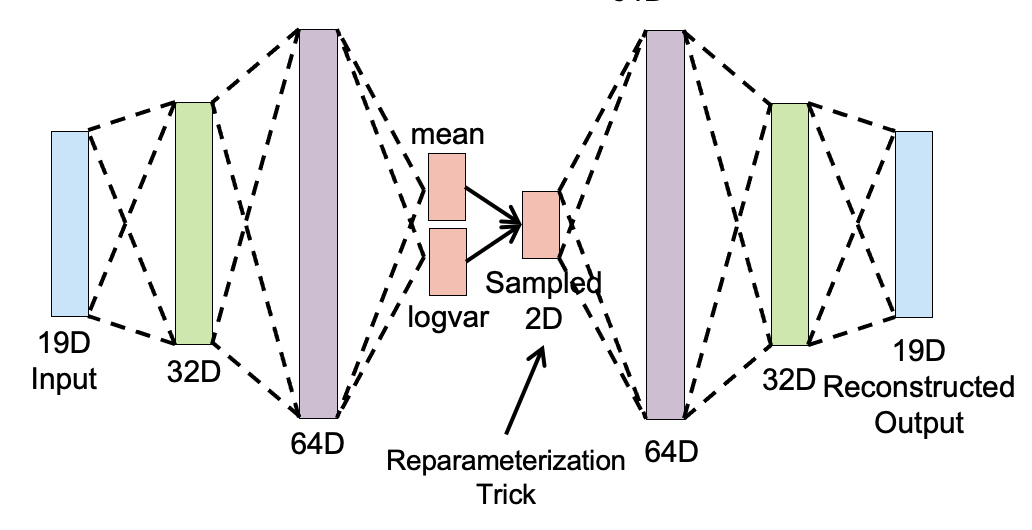}
        \caption{Structure of a VAE with a 19-32-64-2-64-32-19 architecture.}
        \label{fig:VAE_archi}
    \end{subfigure}
    
    \caption{To ensure consistency and facilitate comparative analysis with t-SNE and UMAP, both models utilized their decoder parts to reconstruct the 2-dimensional latent representations back to the original 19-dimensional space. Then we measure MSE and explained variance to evaluate their performances.}
    \label{fig:AEVAE}
\end{figure}

Furthermore, it’s worth noting that other aspects of the Autoencoder and VAE models, such as the optimization algorithm, the number of epochs, the implementation of early stopping, and the use of the MSE loss function, also align with the neural network used for t-SNE and UMAP. The only difference lies in optimizing the total loss during the training process of VAE, which includes MSE and KL divergence terms, but the optimization still focuses on minimizing the MSE on the test set. This consistency ensures a fair comparison between the different dimensionality reduction techniques.

\subsection{Visualization and Comparison}
After dimensionality reduction, we visualized the distribution of objects in the latent space generated by the five techniques, coloring them based on mean metallicity [Fe/H]. We also compared the explained variance through boxplots obtained from the reconstruction by neural networks. Additionally, we selected a sample from the dataset and plotted line charts to compare the reconstructed results of these algorithms with the original data.

\section{Results} \label{sec:Results}
\subsection{2-dimensional Representation of Five Algorithms}
Figure 3 depicts 2-dimensional representations of our 19-dimensional chemical abundances. PCA and Autoencoder exhibit two clusters, while t-SNE and UMAP reveal three clusters, but it’s hard to recognize in VAE representation.

The structure of PCA is connected, since it operates under the assumption of linear relationships, and it seeks to find a linear transformation of the data that captures the most variance. In contrast, UMAP does not rely on linear relationships and can capture more complex, nonlinear relationships in the data. Therefore, it is not typically connected as PCA does. UMAP has the best 2-dimensional visualization as it balanced both local and global structure.

The circular or brain-like shape observed in two-dimensional t-SNE visualizations is attributed to its utilization of t-distribution resembling Gaussian distribution, and this algorithm's tendency emphasizes local structure in the data. t-SNE is designed to preserve local relationships between data points, meaning that nearby points in the high-dimensional space are encouraged to remain close together in the low-dimensional representation. Notably, PCA and UMAP outperform t-SNE in preserving the global structure of the data, resulting in more coherent graphs.

However, the representation of Autoencoder looks weird. This is because we don’t know regularity of the latent space. Specifically, encoding 19-dimensional chemical abundances into a single point in latent space, each unique combination of chemical abundances is mapped to a single point in the latent space. This compression provides a high degree of freedom, meaning that multiple different combinations of chemical abundances can potentially map to the same point in the latent space, leading to overfitting. So, when visualizing the representation generated by the Autoencoder, points in the latent space appears random. Conversely, VAE encodes input data as a distribution approximating Gaussian distribution, manifesting as circular patterns. 

By measuring the strength of absorption lines in stellar spectra, it is possible to infer the abundance of different elements in the stellar atmosphere. Among them, the abundance of iron (Fe) (typically represented by [Fe/H]) is one of the important indicators of interest to astronomers because iron plays a significant role in the formation and evolution processes of stars, and the abundance of iron is closely related to the age and chemical evolutionary history of stars. Consequently, metallicity [Fe/H] serves as a primary contributor to the variation in the data, evident in the trend observed when coloring latent representations.\

\begin{figure}[!ht]
    \centering
    \begin{subfigure}{0.29\textwidth}
        \includegraphics[width=\linewidth]{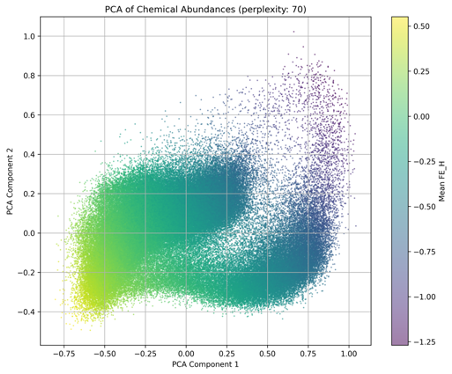}
        \caption{PCA}
        \label{fig:PCA_FE_H}
    \end{subfigure}
    \hfill
    \begin{subfigure}{0.29\textwidth}
        \includegraphics[width=\linewidth]{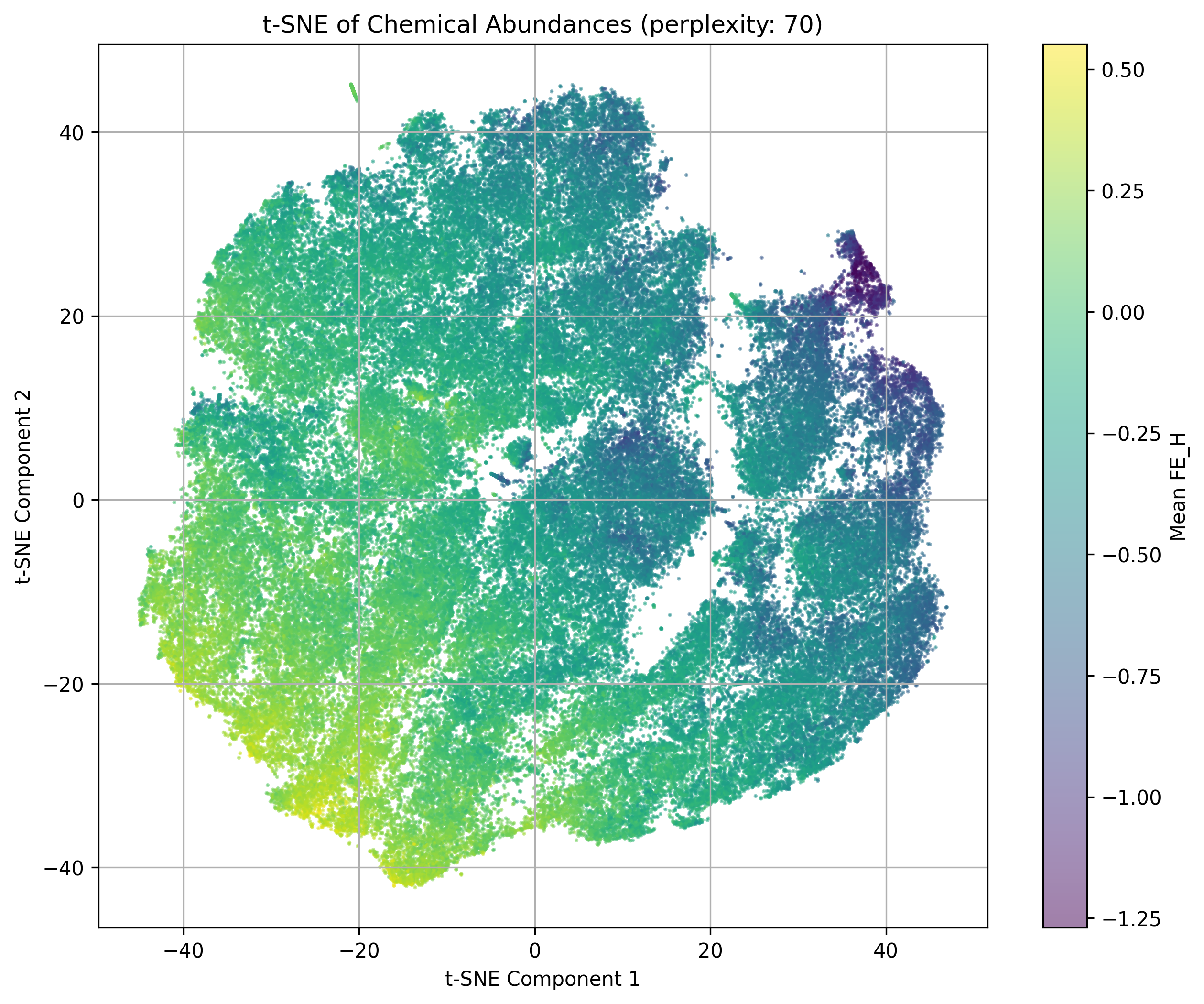}
        \caption{t-SNE}
        \label{fig:tSNE_FE_H}
    \end{subfigure}
    \hfill
    \begin{subfigure}{0.29\textwidth}
        \includegraphics[width=\linewidth]{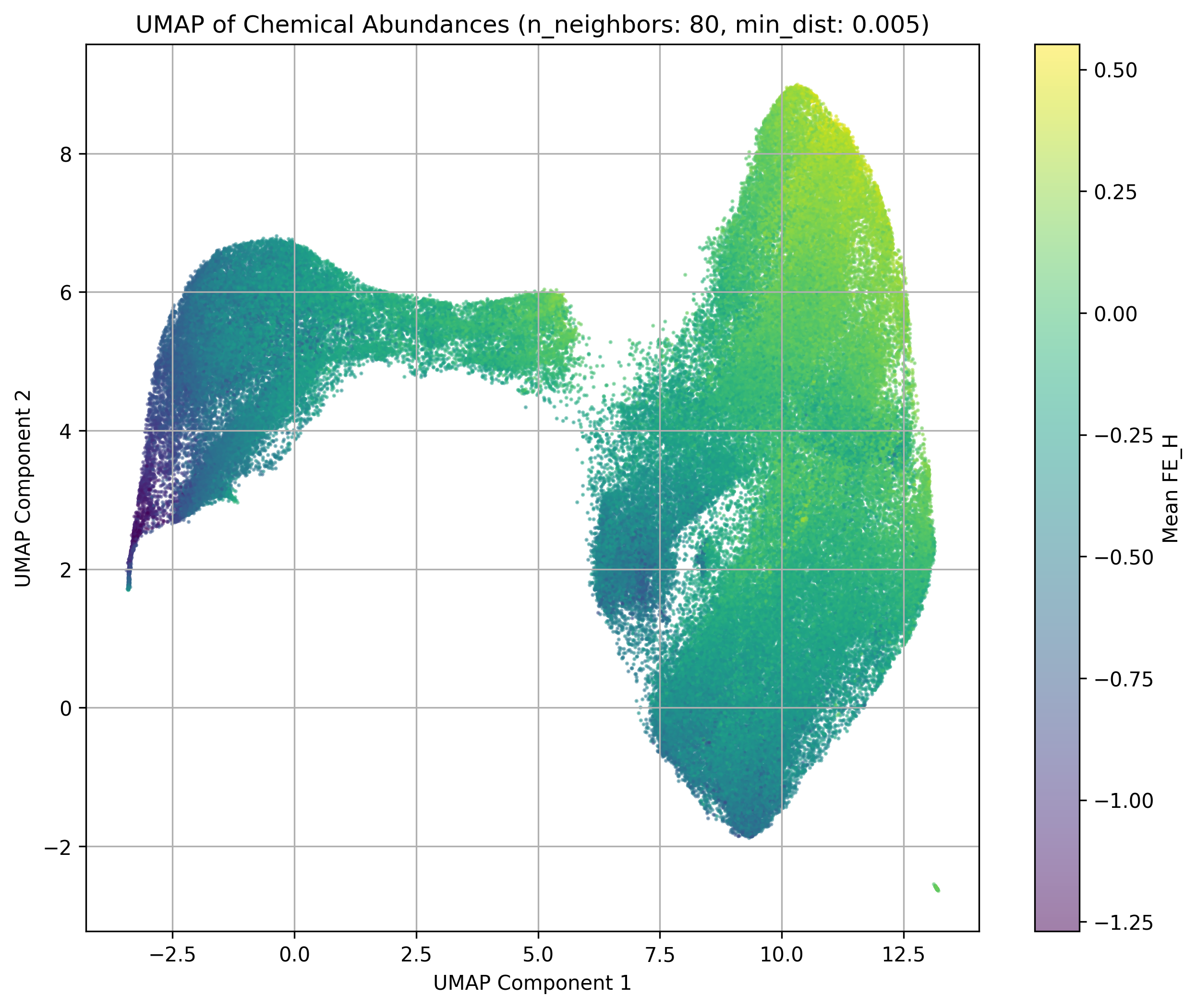}
        \caption{UMAP}
        \label{fig:UMAP_FE_H}
    \end{subfigure}
    
    \medskip

    \centering
    \begin{subfigure}{0.29\textwidth}
        \includegraphics[width=\linewidth]{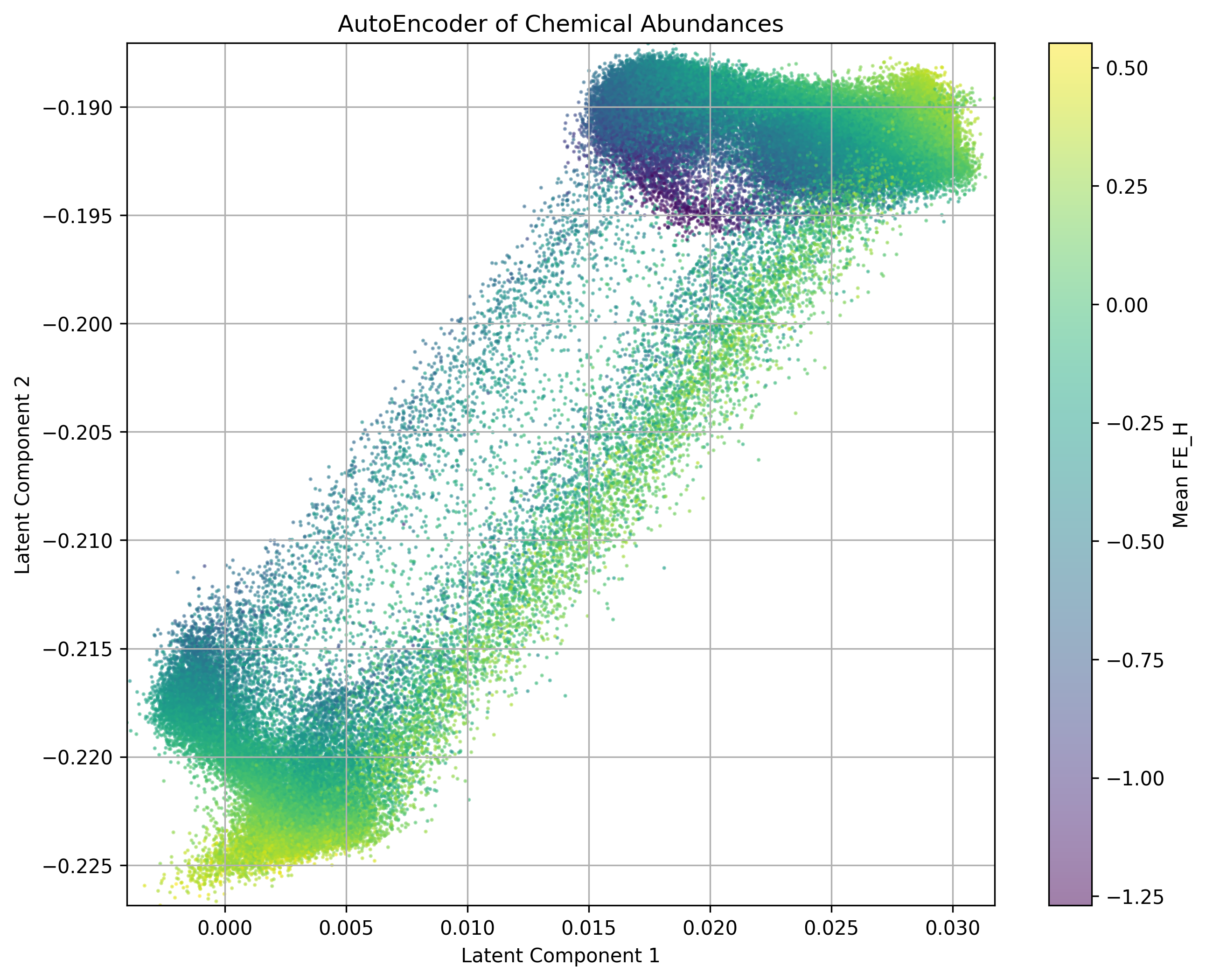}
        \caption{Autoencoder}
        \label{fig:AE_FE_H}
    \end{subfigure}
    \hfill
    \begin{subfigure}{0.29\textwidth}
        \includegraphics[width=\linewidth]{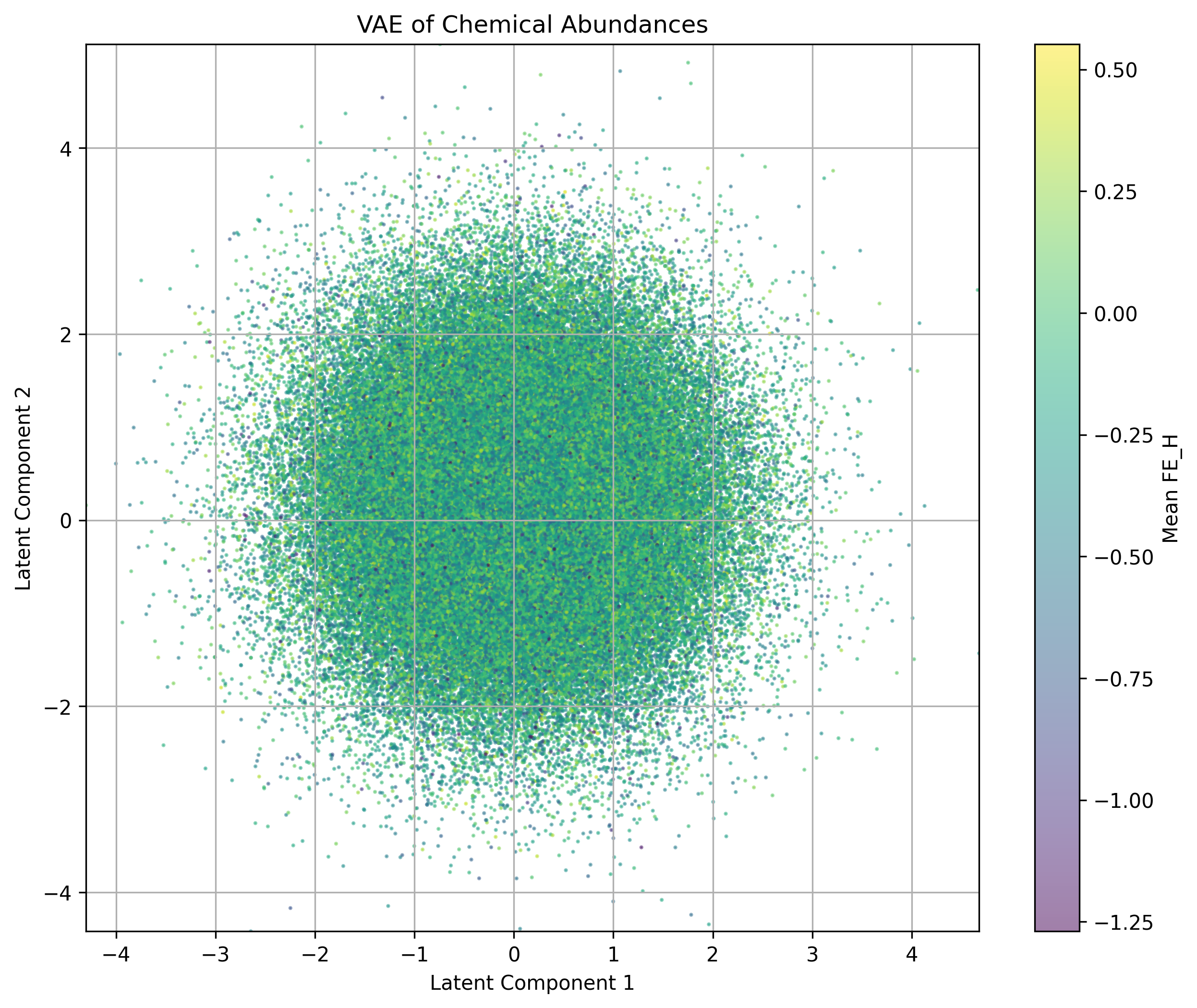}
        \caption{VAE}
        \label{fig:VAE_FE_H}
    \end{subfigure}
    
    \caption{2-dimensional representations for five algorithms}
    \label{fig:2D_repre}
\end{figure}

\subsection{Explained Variance Comparison}
The boxplots in Figure 4 display the performance of these algorithms and Table 1 provides detailed results, including epoch when model converged, optimized MSE, and explained variance. PCA, with its linear assumption, achieves an explained variance of 0.7689. For t-SNE, the explained variance is 0.8535 with an MSE of 0.1465, while UMAP achieves an explained variance of 0.8356 with an MSE of 0.1644. UMAP’s performance is slightly worse than t-SNE, possibly because UMAP preserves more global structure at the expense of explained variance in our case. The Autoencoder achieves an explained variance of 0.8632 with an MSE of 0.1368, while VAE achieves an explained variance of 0.8606 with an MSE of 0.1394. The performances of the Autoencoder and VAE are quite close. VAE incorporates KL divergence in the loss function, but this only increases model stability, with no improvement in explained variance. 

Since we found best MSE for each algorithm, so we can compare their explained variance. The performance ranking of these algorithms is  PCA \textless{} UMAP \textless{} t-SNE \textless{} VAE \textless{} Autoencoder. PCA achieves the lowest explained variance among all algorithms, indicating that it may not capture as much of the variability in the data compared to the other techniques. UMAP and t-SNE perform better than PCA in terms of explained variance, suggesting that they are more effective at preserving the overall structure of the data.Autoencoder and VAE achieve the highest explained variance, indicating that they are able to capture a larger proportion of the variance in the data compared to the other algorithms. The performance gap between linear (PCA) and non-linear (Autoencoder and VAE) algorithms is approximately 10\%, which suggests that non-linear techniques are more effective at capturing the underlying structure and complexity of the data and we can call it as non-linearity gap. Additionally, Figure 5 displays the training and testing loss to demonstrate that our models have converged, which aligns with their performance. 

\begin{figure}[ht]
    \centering
    \includegraphics[width=0.47\linewidth]{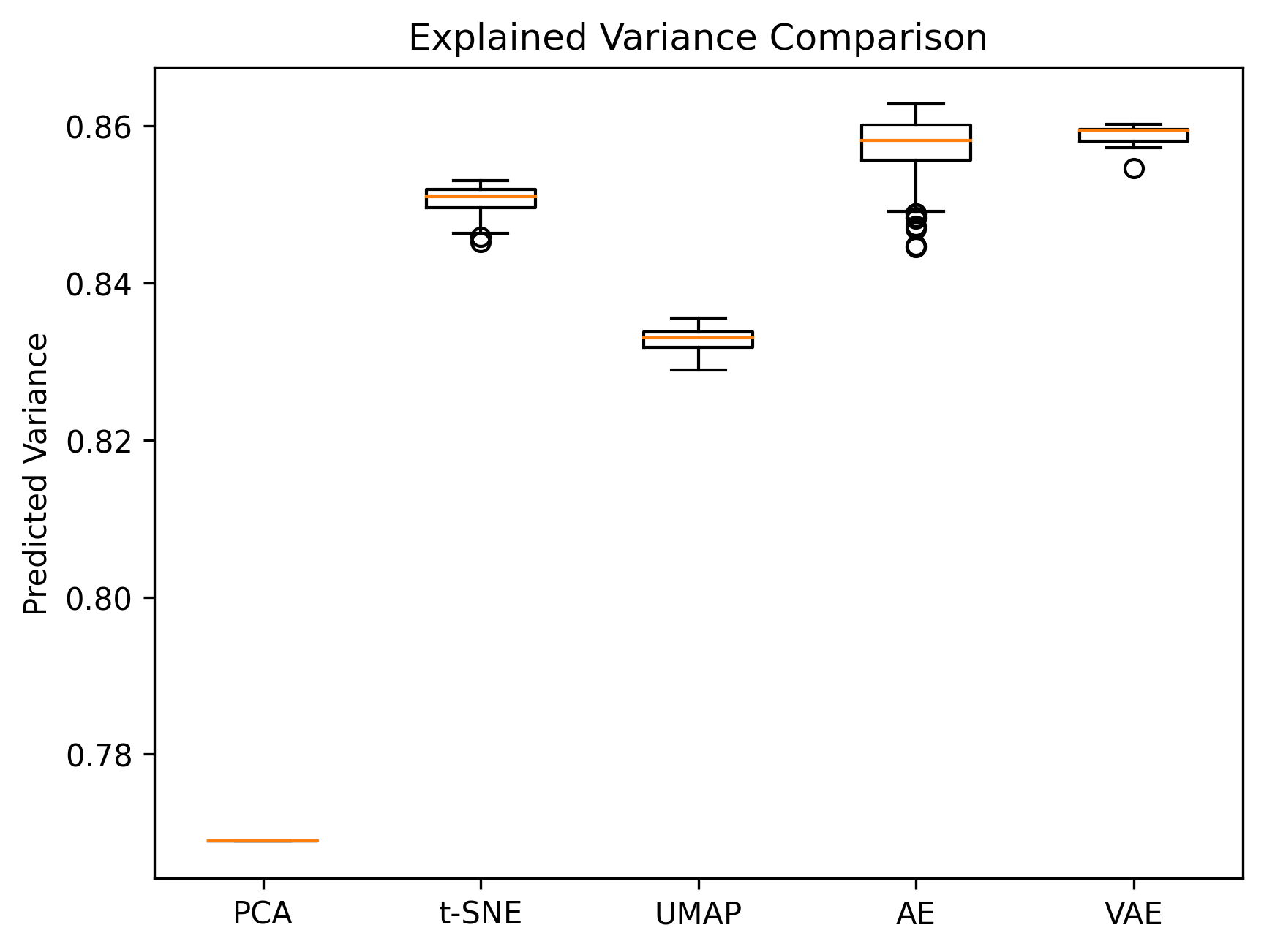}
    \caption{Explained variance comparison for five algorithms}
    \label{fig:ExpVar}
\end{figure}

\begin{table}[ht]
    \centering
    \begin{tabular}{|c|c|c|c|}
        \hline
        Algorithm & Epoch When Converged & Best MSE & Explained Variance \\
        \hline
        PCA & - & - & 0.7689 \\
        \hline
        t-SNE & 2954 & 0.1465 & 0.8535 \\
        \hline
        UMAP & 2931 & 0.1644 & 0.8356 \\
        \hline
        Autoencoder & 2820 & 0.1368 & 0.8632 \\
        \hline
        VAE & 2982 & 0.1394 & 0.8606 \\
        \hline
    \end{tabular}
    \caption{Overview of five algorithms}
    \label{tab: survey_info}
\end{table}

\begin{figure}[ht]
    \centering
    \begin{subfigure}{0.95\textwidth}
        \includegraphics[width=\linewidth]{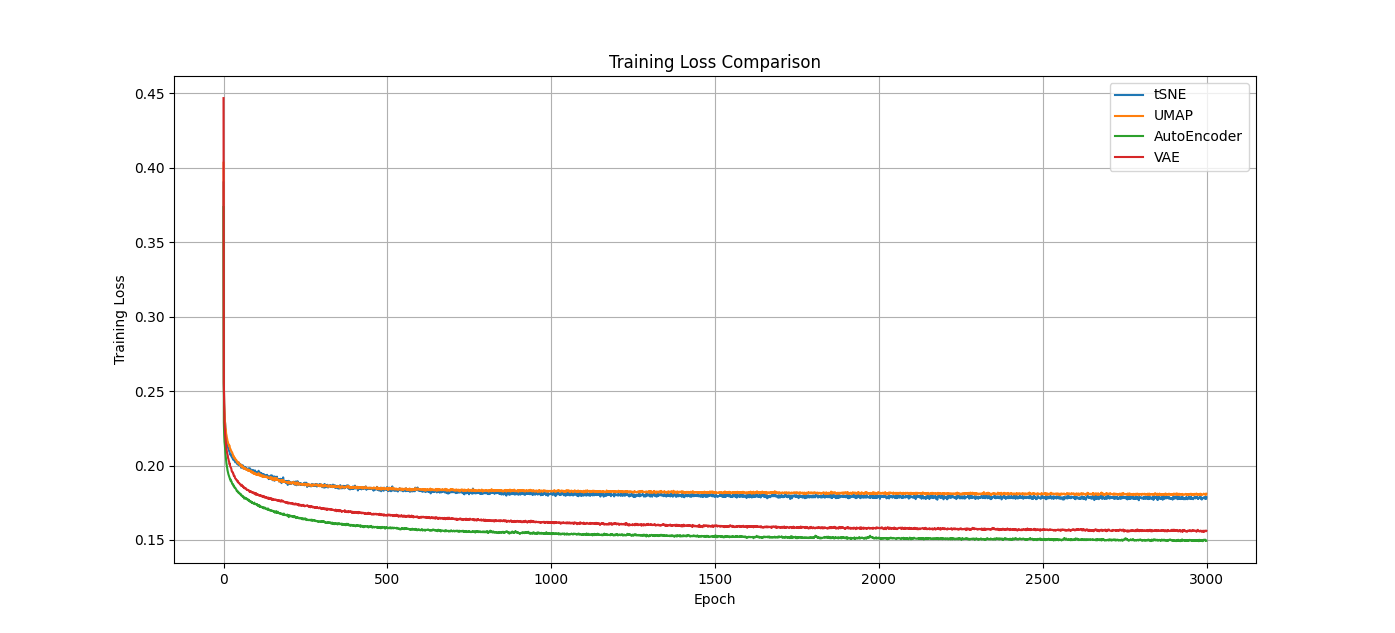}
        \caption{Training Loss Comparison}
        \label{fig:trainloss}
    \end{subfigure}
    
    \medskip
    
    \begin{subfigure}{0.95\textwidth}
        \includegraphics[width=\linewidth]{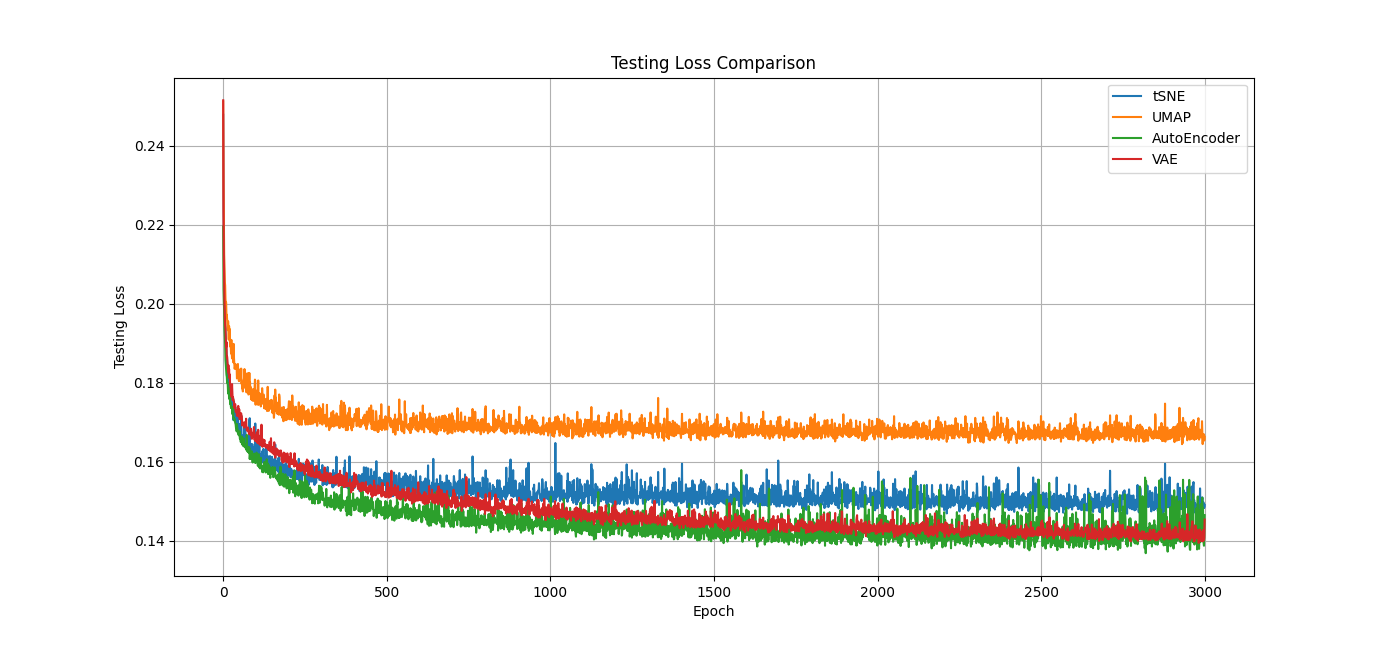}
        \caption{Testing Loss Comparison}
        \label{fig:testloss}
    \end{subfigure}
    
    \caption{Training and testing loss comparison for t-SNE, UMAP, Autoencoder, and VAE.}
    \label{fig:traintestloss}
\end{figure}

\subsection{Reconstruction versus Original data}
Figure 6 illustrates the comparison between the original 19 chemical abundances and the reconstructed outputs from the five algorithms. There is a noticeable gap for each abundance between the PCA reconstructed data and the original data. For t-SNE, the gap gradually narrows down, and some abundances, such as [O/Fe], [S/Fe], and [Ti/Fe], are matched more closely. Similarly, UMAP exhibits a narrowing gap, though discrepancies remain for abundances like [C/Fe], [Cl/Fe], [N/Fe], and [P/Fe]. Autoencoder and VAE show further improvement, with abundances like [C/Fe], [Cl/Fe], [N/Fe], and [P/Fe] becoming even closer to the original data.

\begin{figure}[ht]
    \centering
    \begin{subfigure}{0.32\textwidth}
        \includegraphics[width=\linewidth]{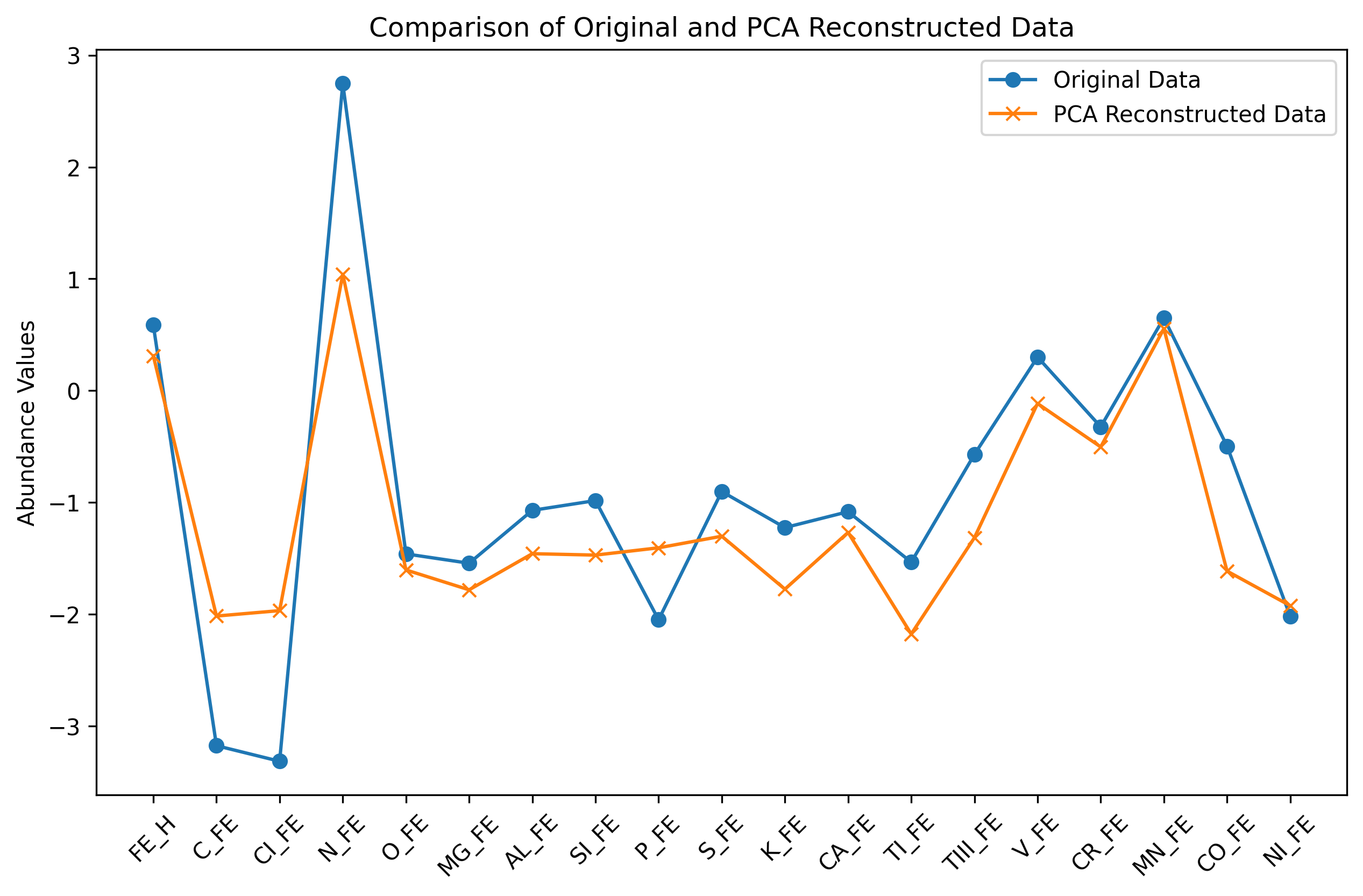}
        \caption{Original vs. PCA Recon}
        \label{fig:test_pca}
    \end{subfigure}
    \hfill
    \begin{subfigure}{0.32\textwidth}
        \includegraphics[width=\linewidth]{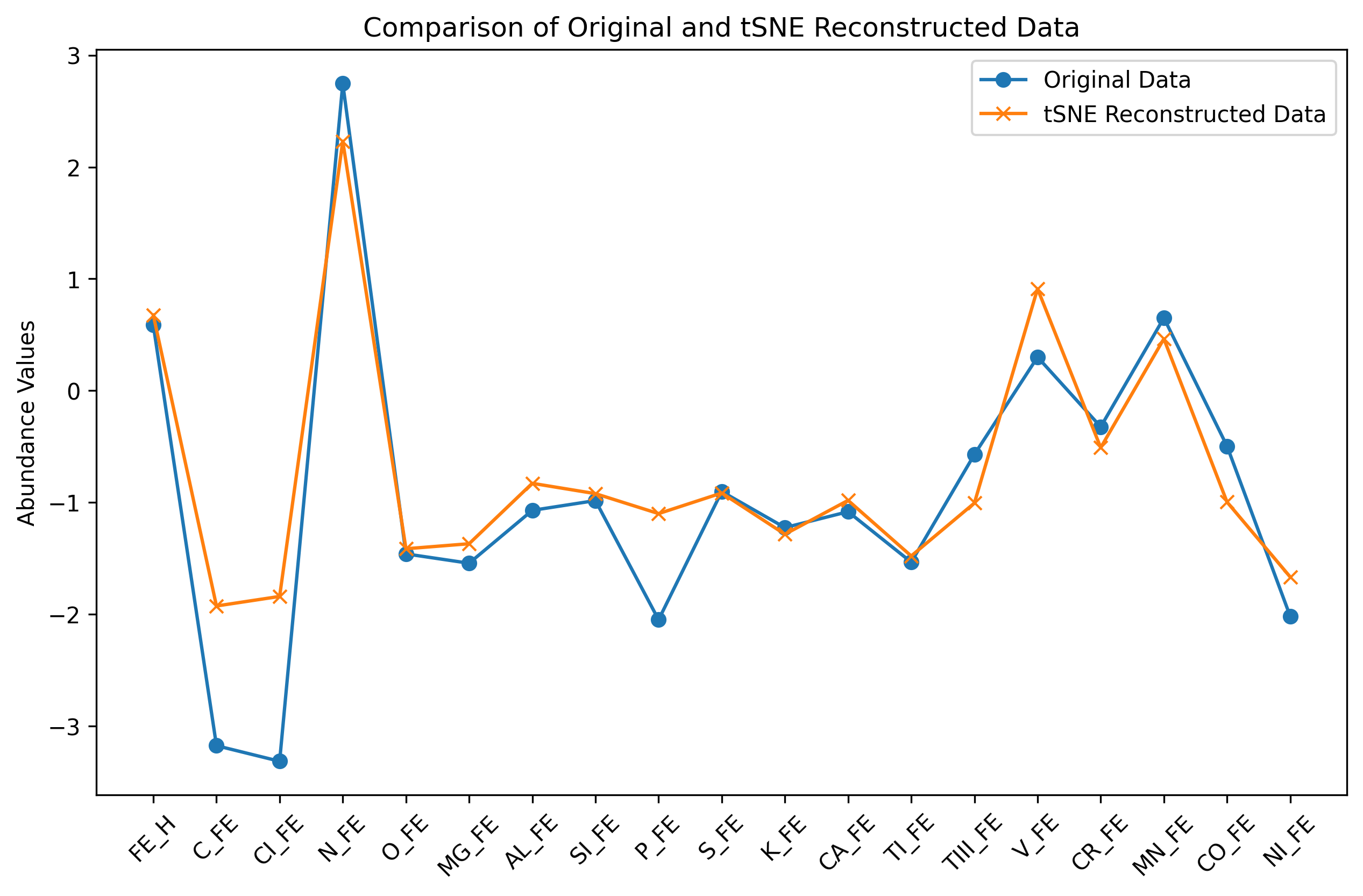}
        \caption{Original vs. t-SNE Recon}
        \label{fig:test_tsne}
    \end{subfigure}
    \hfill
    \begin{subfigure}{0.32\textwidth}
        \includegraphics[width=\linewidth]{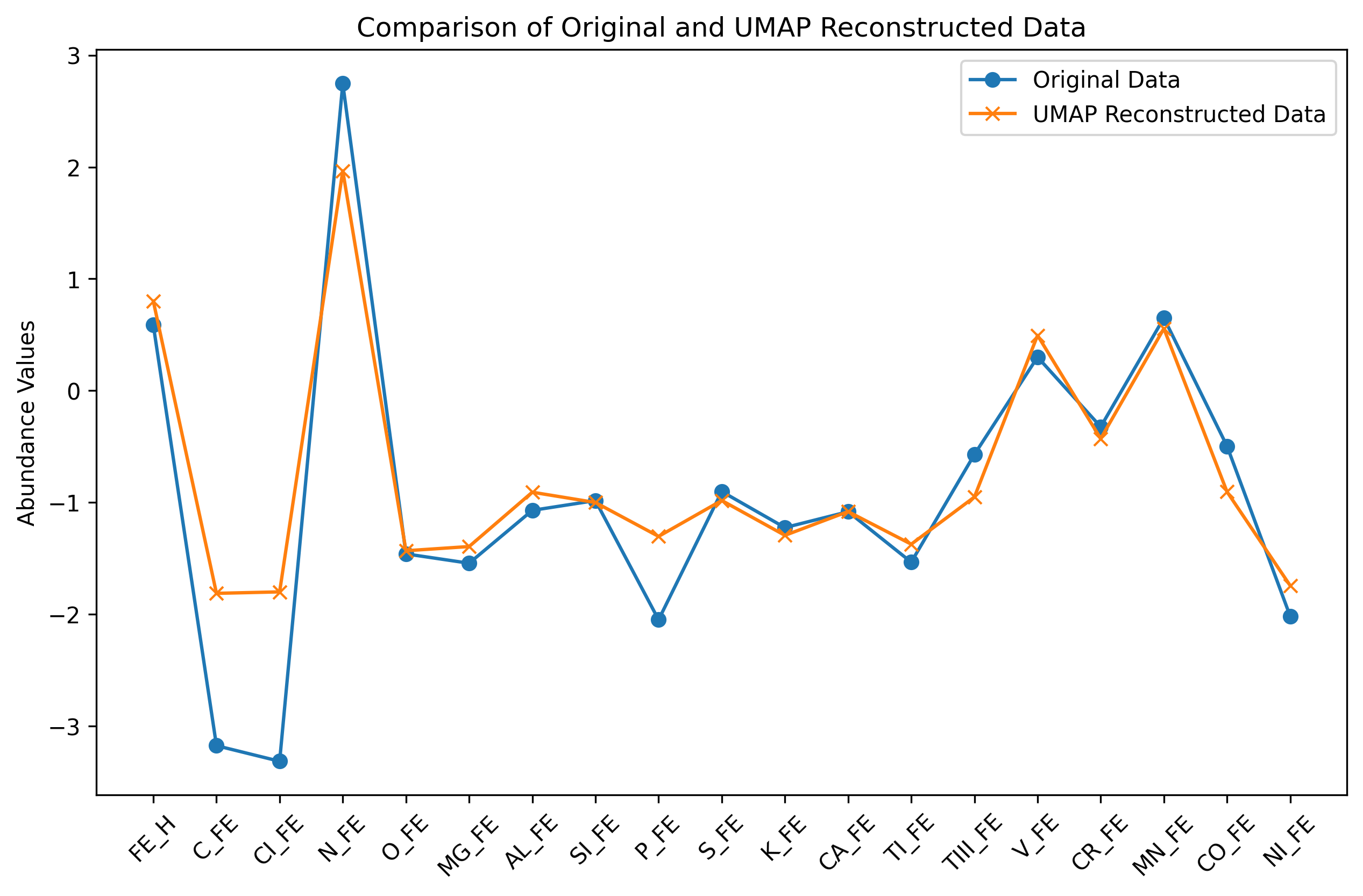}
        \caption{Original vs. UMAP Recon}
        \label{fig:test_umap}
    \end{subfigure}
    
    \medskip

    \centering
    \begin{subfigure}{0.32\textwidth}
        \includegraphics[width=\linewidth]{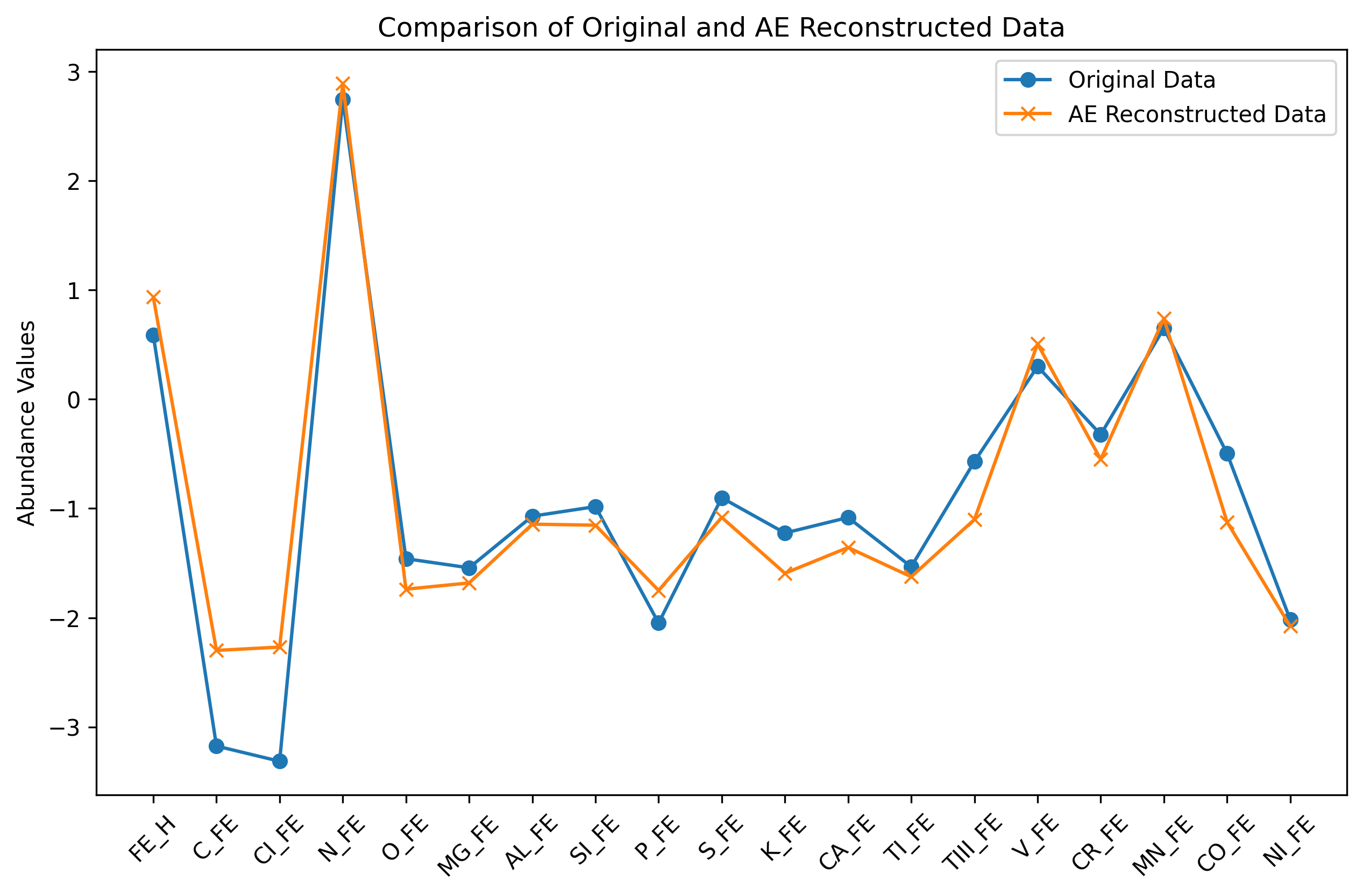}
        \caption{Original vs. Autoencoder Recon}
        \label{fig:test_ae}
    \end{subfigure}
    \hfill
    \begin{subfigure}{0.32\textwidth}
        \includegraphics[width=\linewidth]{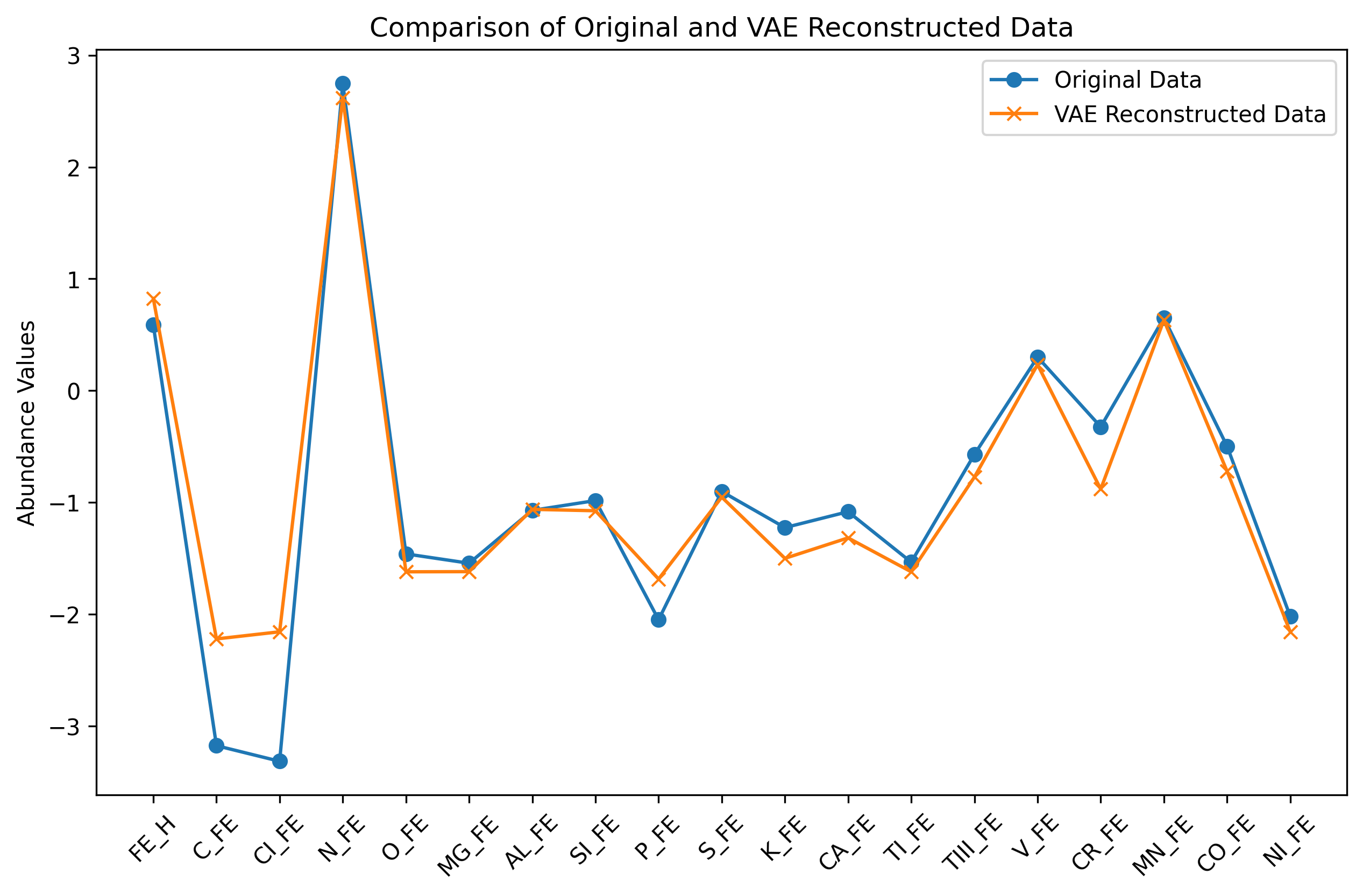}
        \caption{Original vs. VAE Recon}
        \label{fig:test_vae}
    \end{subfigure}
    
    \caption{Comparison between original 19 chemical abundances and reconstructed outputs from the five algorithms.}
    \label{fig:original_vs_recons}
\end{figure}

\section{Discussion}
Although our study explored the latent representations of chemical abundances and evaluated the preservation of information, there still exists limitations will affect results. We only evaluated five dimensionality reduction techniques, and more techniques could be considered in the future work. Also, we only used 2-64-32-19 architecture neural network to predict reconstructed values for t-SNE and UMAP respectively. If we try to utilize other architectures and use different hyperparameters, the explained variance might change as result of optimized MSE changes.

Additionally, we didn’t consider uncertainties in measurements and the potential impact of intrinsic scatter within these relations. So, in the future, we need to investigate ways to incorporate these uncertainties into deep learning-based approaches, such as extension VAEs, which is the variations of the basic VAE framework including modifications to the architecture, loss functions, or training procedures of VAEs.

\section{Conclusion}
Efficiently exploring large astronomical data is an important problem that can be addressed with dimensionality reduction techniques. In this study, we investigated latent representations of chemical abundances using five such techniques and evaluated their performance based on explained variance. Our findings revealed a performance ranking of PCA \textless{} UMAP \textless{} t-SNE \textless{} VAE \textless{} Autoencoder by comparing their explained variance under optimized MSE. The performance of non-linear (Autoencoder and VAE) algorithms has approximately 10\% improvement compared to linear (PCA) algotirhm. This difference can be referred to as the "non-linearity gap." Additionally, we compared reconstructed outputs with the original 19 chemical abundances which provided more explicit differences for each dimensionality reduction techniques Future work should focus on incorporating measurement errors into extension VAEs, thereby enhancing the reliability and interpretability of chemical abundance exploration in astronomical spectra.

\section{Acknowledgement}
I would like to express my sincere appreciation to Prof. Joshua S. Speagle for his guidance and support throughout this research course.

\bibliography{references}{}
\bibliographystyle{aasjournal}

\end{document}